\documentclass[conference]{IEEEtran}
\usepackage{url} 
\usepackage{amsmath, amssymb, amsthm}
\usepackage{hyperref}
\usepackage{graphicx}
\usepackage{gensymb} 
\usepackage[caption=false]{subfig}

\usepackage{float}
\begin{document}

\title{Selection of Most Effective Control Variables for Solving Optimal Power Flow Using Sensitivity Analysis in Particle Swarm Algorithm}

\author{\IEEEauthorblockN{Mohamed Abuella}
\IEEEauthorblockA{Departmtent of Electrical and Computer Engineering\\
University of North Carolina at Charlotte\\
Charlotte, USA\\
Email: mabuella@uncc.edu}
\and
\IEEEauthorblockN{Constantine J. Hatziadoniu}
\IEEEauthorblockA{Department of Electrical and Computer Engineering\\
Southern Illinois University\\
Carbondale, USA\\
Email: hatz@siu.edu }}

\maketitle

\begin{abstract}
Solving the optimal power flow problem is one of the main objectives in electrical power systems analysis and design. The modern optimization algorithms such as the evolutionary algorithms are also adopted to solve this problem, especially when the intermittency nature of generation resources are included, as in wind and solar energy resources, where the models are stochastic and non-linear. This paper uses the particle swarm optimization algorithm for solving the optimal power flow for IEEE-30 bus system. In addition to selection of the most effective control variables based on sensitivity analysis to alleviate the violations and return the system back to its normal state. This adopted strategy would decrease the optimal power flow calculation burden by particle swarm optimization algorithm, especially with large systems.

\indent{{\textbf{\textit{Keywords| Optimal power flow; Particle swarm; Sensitivity analysis.}}}}
\end{abstract}

\section{Introduction} \label{intro}  \label{lit_review}
Optimal power flow (OPF) is a major tool that has been extensively researched since it has introduced. It deals with the minimum cost of power production in electrical power systems with certain constraints \cite{Wood}, \cite{Modern}.
The studies about OPF methods can be traced back to the 1960s when Carpentier and Siroux firstly discussed the OPF problem, and then Domme and Tinney presented a simplified derivative algorithm which is the first practicable algorithm for OPF problem \cite{Wang2005}. But in this algorithm the vibration phenomenon is appearing closing to the optimum. Since then, various kinds of mathematical programming approaches, based on linear and nonlinear programming were proposed in succession, including Newton method, quadratic programming, and interior-point method \cite{Wang2005}.

The aforementioned methods utilize first or second derivative in essence. By this way, it is apt to fall into local optima. Furthermore, there is a difficulty of applying derivative-based optimization techniques to solve OPF for systems that including variable generations. Therefore, various non-classical optimization methods have emerged to cope with some of these shortcomings \cite{Hetzer}. The main modern optimization techniques are genetic algorithm (GA), evolutionary programming (EP), artificial neural network (ANN), simulated annealing (SA), ant colony optimization (ACO), and particle swarm optimization (PSO). They have been successfully applied to wide range of optimization problems in which global solutions are more preferred than local ones \cite{Vlachogiannis2009}, \cite{hwary_enhanced}.

Kennedy and Eberhart introduced PSO as a new heuristic method \cite{Ken}. Different PSO applications in power systems are covered in \cite{hwary_enhanced}. Abido introduced PSO to solve the OPF problem \cite{Abido}. References \cite{Biskas}, \cite{Gaing} have the mechanism of implementing PSO for solving OPF.

The sensitivity analysis of OPF is applied to deduce the sensitivity matrices of the voltage and the current states. The fundamentals of sensitivity analysis of OPF can be found with details in \cite{sensitvity1968}, \cite{Sensitivity1990}, and \cite{Sensitivity2006}. 
The sensitivity analysis is applied in power systems for voltage stability studies as a fast indication of the voltage collapse and a safe voltage margin \cite{tamura1983}, \cite{demarco1990}, \cite{lof1992}, and \cite{gao1992}.

The rest of the paper is organized as follows.
Section~\ref{OPF} gives a description of the optimal power flow problem. Section~\ref{sec:appC} introduces the mathematical derivation of the sensitivity analysis. Section~\ref{PSO} particle swarm optimization (PSO) algorithm is introduced with its parameters and  its mechanism is also explained.
In Section~\ref{simulation} implementation of PSO and the sensitivity analysis to find the OPF solution for several study cases are presented and the discussion of the results is included as well.
Finally, Section~\ref{conclusion} gives the conclusions.


\section{Optimal Power Flow}\label{OPF}
The optimal power flow is an optimization problem to find the optimal allocation of output power among the available generators with given constraints without violations of the operation and the security constraints of the system. The optimal allocation depends on various factors, such as operating cost, system security (or risk) and $CO_2$ emissions, in general they are called cost factors.
The objective function of the optimization problem in this paper is to minimize the operating cost of real power generation, while keeping the electrical system in normal and secure operation condition.

The objective function and the constraints are mostly nonlinear and many methods and algorithms have been developed on the basis of cost factors; generation source type, conventional or renewable; uncertainty treatment (i.e. deterministic or stochastic). For instance, Lagrangian relaxation, direct search method, evolution programming, particle swarm optimization, genetic algorithms, and simulated annealing are some of the solution methods for the optimal power flow problem \cite{Wei}.

In mathematical terms, the optimal power flow (OPF) is an optimization problem set up to minimize an objective function subject to equality and inequality constraints.

The equality constraints are the power flow equations, while the inequality constraints are the limits on control variables and the operating limits of power system dependent (state) variables. The control variables include the generator real power, the generator bus voltages, the transformer tap settings, and the reactive power of switchable VAR sources.
On the other hand, the dependent or state variables include the load bus voltages, the generator reactive powers, and the line flows.
Generally, the OPF problem is a large-scale highly constrained nonlinear non-convex optimization problem.

Mathematically, the OPF problem can be formulated as follow:

\begin{equation}\label{eq objopf}
Min\quad J(\textbf{x,u})
 \end{equation}
 Subject to:
\begin{equation}\label{eq opfeq}
g(\textbf{x,u})=0
 \end{equation}
\begin{equation}\label{eq opfineq}
h(\textbf{x,u}) \leq 0
 \end{equation}

Where \emph{J} is the objective function to be minimized, it could be the cost of real power of thermal units, wind-powered units, or mix of them. \emph{g} is the equality constraints represent the power flow equations. \emph{h} is the inequality constraints that represent the operating limits of the system.\\
 Where $g(\textbf{x,u})=0$ are the balanced power flow equations as following:

\begin{equation}\label{eq PPf}
P_{i}-V_{i}\sum^{n}_{j=1}V_{j}Y_{ij}\cos(\delta_{i}-\delta_{j}-\theta_{ij})=0
\end{equation}
\begin{equation} \label{eq QPf}
Q_{i}-V_{i}\sum^{n}_{j=1}V_{j}Y_{ij}\sin(\delta_{i}-\delta_{j}-\theta_{ij})=0
\end{equation}
Where \emph{$P_i$} is the specified real power at bus \emph{i}, and it equals to the difference between the generation and demand real power (\emph{$P_{Gi}$}-\emph{$P_{Di}$}) at bus \emph{i}, and the same for \emph{$Q_i$}. $Y_{ij}$ is the admittance between buses \emph{i} and \emph{j}. $V_i $ is the voltage magnitude of bus \emph{i} and $\delta_{i}$ is the phase angle of the voltage at bus \emph{i}.

In equations (\ref{eq objopf}-\ref{eq opfineq}), \textbf{x} is the vector of dependent (state) variables consisting of slack bus power $P_{G1}$, load bus voltages $V_L$, generator reactive power output $Q_G$, and transmission  line ratings (loadings) $S_{line}$.
Here x can be expressed as:
\begin{equation} \label{eq x}
\textbf{x}^T=[P_{G_1},\,V_{L_1}...V_{L_{NL}},\,Q_{G_1}...Q_{G_{NG}},\,S_{line_1}... S_{line_{nl}}]
\end{equation}
Where the indices \emph{NL}, \emph{NG}, and \emph{nl} are number of load buses, number of generators, and number of transmission lines respectively.

\textbf{u} is the vector of independent (control) variables consisting of generator voltages \emph{$V_G$}, generator real power outputs $P_G$ except the slack bus \emph{$P_{G1}$}, transformer tap settings \emph{T}, and shunt VAR compensations \emph{$Q_C$}.\\
Hence, \textbf{u} can be expressed as:
\begin{equation} \label{eq u}
\textbf{u}^T=[V_{G_1}...V_{G_{NG}},\,P_{G_2}...P_{G_{NG}},\,T_1...T_{NT},\,Q_{C_1}...Q_{C_{NC}}]
\end{equation}
Where indices \emph{NT} and \emph{NC} are the number of the regulating transformers and shunt compensators respectively.

\begin{itemize}
  \item Generating constraints:\\
  Generator voltages, real power outputs, and reactive power outputs are restricted by their lower and upper limits as follows:
  \begin{equation} \label{eq VG limit}
  V_{G_i}^{min} \leq V_{G_i} \leq V_{G_i}^{max},\qquad i=1,...,NG
  \end {equation}
  \begin{equation} \label{eq PG limit}
  P_{G_i}^{min} \leq P_{G_i} \leq P_{G_i}^{max},\qquad i=1,...,NG
  \end {equation}
  \begin{equation} \label{eq QG limit}
  Q_{G_i}^{min} \leq Q_{G_i} \leq Q_{G_i}^{max},\qquad i=1,...,NG
  \end {equation}
  \item	Transformer constraints:\\
  Transformer tap settings are bounded as follows:
  \begin{equation} \label{eq T limit}
  T_i^{min} \leq T_i \leq T_i^{max},\qquad i=1,...,NT
  \end {equation}
  \item Shunt VAR constraints:\\
  Shunt VAR compensations are restricted by their limits as follows:
  \begin{equation} \label{eq Qc limit}
  Q_{C_i}^{min} \leq Q_{C_i} \leq Q_{C_i}^{max},\qquad i=1,...,NC
  \end {equation}
  \item Security constraints:\\
  These include the constraint of voltages at load buses and the transmission line loadings as follows:
  \begin{equation} \label{eq VL limit}
  V_{L_i}^{min} \leq V_{L_i} \leq V_{L_i}^{max},\qquad i=1,...,NL
  \end {equation}
  \begin{equation} \label{eq Sline limit}
   S_{line_i} \leq S_{line_i}^{max},\qquad i=1,...,nl
  \end {equation}
\end{itemize}

It is worth to mention that the control variables are self-constrained. The hard inequalities of \emph{$P_{G_1}$}, $V_L$, $Q_G$, and $S_{line}$ can be incorporated in the objective function as quadratic penalty terms (penalty functions). Therefore, the objective function in equation (\ref{eq objopf}) can be augmented as follows:
  \begin{equation} \label{eq opfaugj}
  \begin{split}
  J_{aug}=J+\lambda_P(P_{G_1}-P_{G_1}^{lim})^2+\lambda_V\sum^{NL}_{i=1}(V_{L_i}-V_{L_i}^{lim})^2 \\
  +\lambda_Q\sum^{NG}_{i=1}(Q_{G_i}-Q_{G_i}^{lim})^2+\lambda_S\sum^{nl}_{i=1}(S_{line_i}-S_{line_i}^{max})^2
  \end{split}
   \end {equation}
Where $\lambda_P$, $\lambda_V$, $\lambda_Q$, and $\lambda_S$ are penalty factors and $x^{lim}$ is the limit value of the dependent variable \emph{x} given as:
\begin{equation}\label{eq limw}
 x^{lim} = \left\{
  \begin{array}{l l}
    x^{max}; & \quad (x > x^{max})\\
    x^{min}; & \quad (x < x^{min})\\
  \end{array} \right.
\end{equation}
\

\section{Sensitivity Analysis for Optimal Power Flow} \label{sec:appC}
Earlier research on the application of sensitivity analysis in power system belongs to Peschon et al \cite{sensitvity1968}. They introduced two methods. First one can be applicable to normal power flow problems for small changes in the variables such as active generation, and the second method considers the minimization of objective function satisfying some constraints such as power flow equation. Similaresearch was carried out by Gribik et al \cite{Sensitivity1990}.

\subsection{Mathematical Formulation}
The method of calculating the sensitivities of voltages and currents are determined simultaneously, which
are further used to determine the changes in power flows \cite{Sensitivity2006}.
Considering the generalized equations of the form:
\begin{equation}\label{eq sen pf}
g(\textbf{x,u,p}) = 0
\end{equation}
where \emph{g} is 2\emph{N} dimensional vector, and \emph{N} is number of buses.
The variables mentioned in equation (\ref{eq sen pf}) can be categorized as:\\
(\textbf{x}) are dependent (state) variables, these are the controlled variables and they are unknown. \textbf{x} is a 2\emph{N} dimensional vector.\\
(\textbf{u}) are independent control variables, these are the operating variables or imposed variables of the system. \textbf{u} is an \emph{M} dimensional vector.\\
(\textbf{p}) are parameter variables, these are uncontrollable variables and are normally specified in the power flow problem such as the admittance and the loads.

Depending on the variables to be determined, the variables in the power flow problem can be selected as \textbf{x, u,} and \textbf{p}. One might be interested in controlling \emph{M} variables out of the 2\emph{N} variables.\\
If \textbf{x$_0$}, \textbf{u$_0$}, and \textbf{p$_0$} are the initial state vectors, rewriting equation (\ref{eq sen pf}) as:
\begin{equation}\label{eq sen pf0}
g(\textbf{x}_0,\textbf{u}_0,\textbf{p}_0) = 0
\end{equation}
The changes $\Delta\textbf{x}$ corresponding to small changes $\Delta\textbf{u}$ and $\Delta\textbf{p}$, will satisfy the new equations:
\begin{equation}\label{eq sen dpf0}
g(\textbf{x}_0+\Delta\textbf{x}, \textbf{u}_0+\Delta\textbf{u}, \textbf{p}_0+\Delta\textbf{p}) = 0
\end{equation}
Expanding (\ref{eq sen dpf0}) by Taylor's series and neglecting higher order terms,
{\small 
\begin{equation}\label{eq sen pftaylor}
g(\textbf{x}_0+\Delta\textbf{x}, \textbf{u}_0+\Delta\textbf{u}, \textbf{p}_0+\Delta\textbf{p}) =
g(\textbf{x}_0,\textbf{u}_0,\textbf{p}_0)+g_x\Delta\textbf{x}+g_u\Delta\textbf{u}+g_p\Delta\textbf{p}
\end{equation}}
where, $g_x$, $g_u$ and $g_p$ are the partial derivatives of \emph{g} with respect to \emph{x}, \emph{u} and \emph{p} respectively and are given by:

\begin{equation}\label{eq sen gx}
g_x=\frac{\partial(g_1, g_2,..., g_{2N})}{\partial(x_1, x_2,..., x_{2N})}
\end{equation}
where $x_1$, $x_2$,...,$x_{2N}$ are the elements of \textbf{x}.

\begin{equation}\label{eq sen gu}
g_u=\frac{\partial(g_1, g_2,..., g_{2N})}{\partial(u_1, u_2,..., u_{M})}
\end{equation}
where $u_1$, $u_2$,...,$u_M$ are the elements of \textbf{u}.

\begin{equation}\label{eq sen gp}
g_p=\frac{\partial(g_1, g_2,..., g_{2N})}{\partial(p_1, p_2,..., p_{2N})}
\end{equation}
where $p_1$, $p_2$,...,$p_{2N}$ are the elements of \textbf{p}.\\
When changes are small, solution for $\Delta\textbf{x}$ will be:

\begin{equation}\label{eq sen susp}
\Delta\textbf{x}=\textbf{S}_u\Delta\textbf{u}+\textbf{S}_p\Delta\textbf{p}
\end{equation}
where $\textbf{S}_u$ and $\textbf{S}_p$ are the sensitivities of \textbf{x} with respect to \textbf{u} and \textbf{p} respectively and are obtained as:
\begin{equation}\label{eq sen su}
\textbf{S}_u=-{g_x}^{-1}{g_u}
\end{equation}
\begin{equation}\label{eq sen sp}
\textbf{S}_p=-{g_x}^{-1}{g_p}
\end{equation}
If \textbf{p} variables are not changed then (\ref{eq sen susp}) can be rewritten as:
\begin{equation}\label{eq sen sudu}
\Delta\textbf{x}=\textbf{S}_u\Delta\textbf{u}
\end{equation}

The set of dependent and independent variables can be chosen as the system requirements and the problem formulation. Some of the parameters of a type may belong to the set of dependent whereas remaining parameters of same type may belong to the set of independent variables. for instance, as bus voltages they might be considered as independent variables when they are at generator buses while they are considered dependent at load buses.

\subsection{Determination of Voltage Sensitivities at Buses}
\label{sec:voltsens}
Power flow equations are comprising of 6 variables namely \emph{P}, \emph{Q}, \emph{V}, $\delta$ ,\emph{Y} and $\theta$. All the variables can be assumed to be obtained or specified at the base condition. The variables \emph{Y} and $\theta$ are normally specified and are constant. The other variables are not always constant and they are either specified or determined, depending upon the type of buses. The variables for which changes are specified are grouped as independent variables and the variables which are determined against these changes are grouped as dependent variables \cite{Sensitivity2006}.

For the slack bus, \emph{V} and $\delta$ are specified and \emph{P} and \emph{Q} are subjected to change. For generator bus, \emph{P} and \emph{V} are specified and \emph{Q} and $\delta$ are subjected to change. For load buses, \emph{P} and \emph{Q} are specified and \emph{V} and $\delta$ are changed.
Now consider the power system of \emph{N} buses and \emph{B} branches.
Power flow equations can be described by (\ref{eq PPf}, \ref{eq QPf}).
There are 2\emph{N} set of equations and a set of 2\emph{N} variables can be selected as state variables (\textbf{x}) and remaining as control variables (\textbf{u}).

Consider that only \emph{M} control variables are changed and for these changes, it is desired to obtain the changes in the real and reactive power at slack buses, reactive power and angles at generator buses and voltages and angles at load buses. Then, the power flow equations can be written as following:
\begin{equation}\label{eq grPf}
g(V_i,V_j,\delta_i,\delta_j,P_i, Q_i, Y_{ij}, \theta_{ij}) = 0
\end{equation}
Let
\begin{center}
$P_{sl}, P_G, P_L \,\in\,P_i$\\
$Q_{sl}, Q_G, Q_L \,\in\,Q_i$\\
$V_{sl}, V_G, V_L \,\in\,V_i$\\
\end{center}
Grouping the variables of (\ref{eq grPf}) as:
\begin{equation}\label{eq grx}
\textbf{x}=[P_{sl}, Q_{sl}, Q_G, \delta_G,  V_L, \delta_L]
\end {equation}
\begin{equation}\label{eq gru}
\textbf{u}=[V_{sl}, \delta_{sl}, P_G, V_G, P_L, Q_L]
\end {equation}
\begin{equation}\label{eq grp}
\textbf{p}=Y_{ij},\:\theta_{ij}
\end {equation}
From (\ref{eq sen sudu}), the changes in dependent variables can be obtained

\begin{equation}\label{eq supf}
\begin{split}
[\Delta P_{sl}, \Delta Q_{sl}, \Delta Q_G, \Delta \delta_G,  \Delta V_L, \Delta \delta_L] = \\
 \textbf{S}[\Delta V_{sl}, \Delta \delta_{sl}, \Delta P_G, \Delta V_G, \Delta P_L, \Delta Q_L]
 \end{split}
\end{equation}
where \textbf{S} is the sensitivity matrix of order 2\emph{N}x2\emph{N} and can be obtained as given by (\ref{eq sen su}). For slack bus and generator buses following substitution can be made in (\ref{eq supf}):
\begin{equation}\label{eq subst}
\Delta V_{sl}=\Delta V_{G}=\Delta \delta_{sl}=0
\end{equation}

After determining the changes in the load bus voltages, load bus angles and generator bus angles from (\ref{eq supf}) and with the substitutions from (\ref{eq subst}) all the bus voltages and angles can be arranged as:
 \begin{equation}\label{eq sensvolt}
[\Delta \textbf{V}, \Delta \boldsymbol{\delta}] =[\Delta V_{sl}, \Delta V_G, \Delta V_L, \Delta \delta_{sl}, \Delta \delta_G, \Delta \delta_L]
\end{equation}

\subsection{Determination of Current Sensitivities in the Lines}
\label{sec:cursens}
It is well known that the changes in voltage angles and voltage magnitudes will cause changes in branch currents \cite{Sensitivity2006}. These currents in complex form can be expressed as:
\begin{equation}\label{eq current}
I_{ij}=Y_{ij}[V_i(cos \delta_i + j sin \delta_i)-V_j(cos \delta_j + j sin \delta_j)]
\end{equation}
Where $Y_{ij}=|Y_{ij}|\angle \theta_{ij}$ and $I_{ij} \in B$, since B is the number of branches.
Equation (\ref{eq current}) can be written in the form:

\begin{equation}\label{eq grcrnt}
g_{ij}(I_{ij}, Y_{ij}, \theta_{ij}, V_i, V_j, \delta_i, \delta_j) = 0
\end{equation}
Grouping the variables of (\ref{eq grcrnt}) as\\
$\textbf{x}=I_{ij}$\\
$\textbf{u}= V_i, V_j, \delta_i, \delta_j$ (i.e. V and $\delta$ variables at all buses)\\
$\textbf{p}=|Y_{ij}|,\: \theta_{ij}$.\\
Sensitivities of $I_{ij}$ for the changes in $V_i, V_j, \delta_i , \delta_j$ can be obtained from (\ref{eq sen sudu}) as:

\begin{equation}\label{eq d_current}
\Delta I_{ij}=\textbf{R}[\Delta \textbf{V}, \Delta \boldsymbol{\delta}]
\end{equation}
where \textbf{R} is sensitivity matrix obtained by (\ref{eq sen su}) which is given as:
\begin{equation}\label{eq sencrnt}
\textbf{R}=-g_{ijx}^{-1}\:g_{iju}
\end{equation}
With $g_{ijx}$ is Jacobian of $g_{ij}$ with respect to \textbf{x} (i.e. $I_{ij}$).\\
While $g_{iju}$ is Jacobian of $g_{ij}$ with respect to \textbf{u} (i.e. $V_i, V_j, \delta_i , \delta_j$) Substituting from (\ref{eq sensvolt})  and (\ref{eq sencrnt}), the equation (\ref{eq d_current}) can be rewritten as:
{\small 
\begin{equation}\label{eq sencrntfinal}
[\Delta I_{ij}]_{B \times 1}=[\textbf{R}]_{B \times 2N}[\Delta V_{sl}, \Delta V_G, \Delta V_{L} \Delta \delta_{sl}, \Delta \delta_G, \Delta \delta_{L}]_{2N \times 1}
\end{equation}}
Where $ \Delta V_{sl}=\Delta V_{G}=\Delta \delta_{sl}=0$.

\section{Particle Swarm Optimization Algorithm}  \label{PSO}
The original PSO suggested by Kennedy and Eberhart is based on the analogy of swarm of bird and school of fish \cite{Ken}. The algorithm was simplified and it was observed to be performing a solution to an optimization problem.
\subsection{Standard Algorithm}
PSO, as an optimization tool, provides a swarm-based search procedure in which particles change their positions with time. In a PSO system, particles fly around in a multidimensional search space. During flight, each particle adjusts its position according to its own experience, and the experience of neighboring particles, making use of the best position encountered by itself and its neighbors.
When improved positions are being discovered these will then come to guide the movements of the swarm. The process is repeated and by doing so it is hoped, but not guaranteed, that a satisfactory solution will eventually be discovered \cite{Gaing}.

The following is the conventional terminology of the parameters in PSO:
Let \emph{x} and \emph{v} denote a particle coordinates (position) and its corresponding flight speed (velocity) in a search space, respectively. Therefore, the\textbf{ \emph{i}}th particle is represented as $x_i=[x_{i1}, x_{i2},....,x_{im}]$. Since \textbf{\emph{m}} is the last dimension or coordinate of the position of the the\textbf{ \emph{i}}th particle in the search space and so that $\textbf{\emph{d}}=1,2,...,{\emph{m}}$.\\
The best previous position of the\textbf{ \emph{i}}th particle is recorded and represented as, \\ $pbest_i=[pbest_{i1}, pbest_{i2},....,pbest_{im}]$.\\
The position of the best particle among all the particles in the group is represented by the $gbest$.
In a particular dimension \textbf{\emph{d}} there is a group best position which is $gbest_d$.\\
The velocity for the\textbf{ \emph{i}}th particle is represented as,\\ $v_i=[v_{i1}, v_{i2},....,v_{id}]$.
The modified velocity and position of each particle can be calculated by using the following formulas:
\begin{equation} \label{eq pso v}
v_{id}^{k+1}=w*v_{id}^{k}+c_1*U*(pbest_{id}^k-x_{id}^k)+c_2*U*(gbest_{d}^k-x_{id}^k)
\end{equation}
\begin{equation} \label{eq pso x}
x_{id}^{k+1}=x_{id}^k+v_{id}^{k+1}
\end{equation}
$i=1,2,....,n; \qquad d=1,2,...,m$\\
Where\\
$n$ \qquad  number of particles in a group;\\
$m$ \qquad number of members in a particle;\\
$k$ \qquad  pointer of iterations (generations);\\
$w$ \qquad inertia weight factor;\\
$c_1,c_2$ \quad acceleration factors;\\
$U$ \qquad uniform random number in the range [0,1];\\
$x_{id}^k,\! v_{id}^k$ the position and velocity of the\textbf{ \emph{i}}th particle in the\textbf{ \emph{d}}th dimension at iteration k;

\begin{figure}[!ht]
\begin{center}
\includegraphics[width=1.35in]{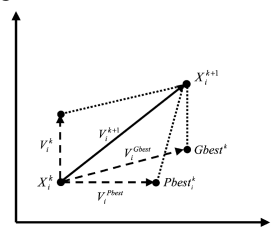}
\caption[PSO search mechanism]{PSO search mechanism \label{fig:pso m}}
\end{center}
\end{figure}

The search mechanism of the PSO using the modified velocity and position of individual based on (\ref{eq pso v}) and (\ref{eq pso x}) is illustrated in Fig. \ref{fig:pso m}.

In the above procedures, the velocity should between $v_d^{min} \leq v_{id} \leq v_d^{max}$ If $v_d^{max}$
is too high, particles might fly past good solutions. If $v_d^{max}$ is too small, particles may not explore sufficiently beyond local solutions. In many experiences with PSO, it is often set at 10 - 20\% of the dynamic range of the variable on each dimension \cite{Gaing}.

The constants $c_1$ and $c_2$ represent the weighting of the stochastic acceleration terms that pull each particle toward the $pbest$ and $gbest$ positions. Low values allow particles to move
far from the target regions before being dragged back. On the other hand, high values result in sudden movement toward, or past, target regions. Hence, the acceleration constants $c_1$ and $c_2$ are often set to be 2 according to empirical experience \cite{Gaing}.

Suitable selection of inertia weight $w$ in (\ref{eq pso v}) provides a balance between global and local explorations, to find a sufficiently optimal solution. As originally developed $w$, often decreases linearly from about 0.9 to 0.4 during the run. In general, the inertia weight is set according to the following equation:
\begin{equation} \label{eq pso x}
w=w_{max}-\frac{(w_{max}-w_{min})}{iter_{max}}\times iter
\end{equation}
Where $iter_{max}$ is the maximum number of iterations (generations), and $iter$ is the current number of iterations.
\begin{figure}[!ht]
\begin{center}
\includegraphics[width=1.3in]{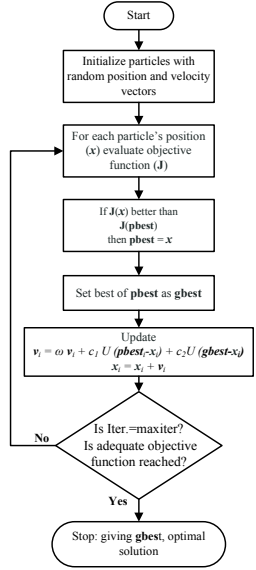}
\caption[PSO algorithm flowchart]{PSO algorithm flowchart\label{fig:psofchart}}
\end{center}
\end{figure}

\subsection{Implementation of PSO for OPF Problem}\label{sec:PSOimplement}
The primary objective of the OPF problem is to determine the optimal combination of power outputs of all generating units so as to meet the required load demand at minimum operating cost while satisfying system equality and inequality constraints. \\
The formulation of OPF for applying PSO is done by separating the problem variables to state variables, \emph{x}, and control variables, \emph{u}, as it was described in equations (\ref{eq objopf}, \ref{eq opfeq}, and \ref{eq opfineq}), they also mentioned here as follows:\\
\begin{equation}\label{eq psoobj}
Min\quad J(\textbf{x,u})
\end{equation}
subject to:
\begin{equation}\label{eq psoeq}
g(\textbf{x,u})=0
\end{equation}
\begin{equation}\label{eq psoineq}
h(\textbf{x,u}) \leq 0
\end{equation}
\begin{equation}\label{eq psour}
\textbf{u} \in U
\end{equation}
Where:
\begin{equation}\label{eq psox}
\textbf{x}=[P_{G_1}, V_L, Q_{G}, S_{line}]
\end{equation}
\begin{equation}\label{eq psou}
\textbf{u}=[P_{G}, V_{G},T, Q_{C}]
\end{equation}
The equality constraint in (\ref{eq psoeq}) are the nonlinear power flow equations as in (\ref{eq PPf} and \ref{eq QPf})\\
The inequality constraints (\ref{eq psoineq}) are the functional operating constraints, such as transmission line limits, load bus voltage magnitude limits, generator reactive capabilities, and slack bus active power output limits. Constraints (\ref{eq psour}) define the feasibility region of the control variables of the problem such as the active power output limits of the generators (except the generator at the slack bus), generation bus voltage magnitude limits, transformer-tap setting limits, and bus shunt admittance limits.

 Each particle in PSO is a vector containing the control variables \textbf{u}, suggesting a possible solution to the OPF problem. Then the position of the \textbf{\emph{i}}th particle $x_i$ can be represented as $x_i = \textbf{u}_i = (u_{i1},u_{i2},..., u_{im})$, where \emph{m} is the number of dimensions and it is also represented the number of control variables, $d\in[1,\,m]$, $u_{id} \in [u_{id}^{min},\, u_{id}^{max}]$.\\
$u_{id}^{min}$, and $u_{id}^{max}$ are the lower and upper bounds of $u_{id}$. The particles are moving in an \emph{m} dimensional space.\\ For consistency's sake, the general definition of the swarm particle is which used in the rest of the paper, as in equations (\ref{eq pso v} and \ref{eq pso x}). Therefore, the\textbf{ \emph{i}}th particle is represented as $x_i=[x_{i1}, x_{i2},....,x_{im}]$ instead of $\textbf{u}_i$.\\
Each particle attempts to minimize the following objective function:

\begin{equation}\label{eq psoobjunarg}
J_{aug}=\sum^{NG}_{i=1}F_i(P_{Gi})+\lambda\left[\sum^{NS}_{i=1}\mu_i*h_i(\textbf{x,u})\right]
\end{equation}
Since:
\begin{equation}\label{eq mulim}
 \mu_i = \left\{
  \begin{array}{l l}
    1; & \quad h_i(\textbf{x,u}) > 0\\
    0; & \quad h_i(\textbf{x,u}) \leq 0\\
  \end{array} \right.
\end{equation}
Here the objective function becomes unconstrained or augmented objective function by using the classical penalty functions principle.
All inequality constraints in equation (\ref{eq psoineq}) replaced by penalty terms. While the power balance equations (\ref{eq psoeq}), which are the equality constraints, is solved for each particle and in every iteration by Newton-Raphson power flow algorithm, therefore no need to use a penalty function for this equality constraint in equation (\ref{eq psoobjunarg}).

$J_{arg}$ is the penalized objective function and $F_i(P_{Gi})$ is the cost of the real power from the generator $P_{Gi}$ while $\lambda$ is the penalty factor for operating constraints. $\mu_i$ is an indicator of occurring any violations and work outside the feasibility region of the solution. It has only two values as in equation (\ref{eq mulim}), it is either 1 when a violation of the limits occurred in the corresponding constraint or 0 when there is no violation. The penalty terms that have been used are quadratic penalty functions as those in equation (\ref{eq opfaugj}). Whereas $NG$ is the number or generators while $NS$ is the number of the state variables to be bounded within their limits.
The penalty factor $\lambda$ is used to penalize the cost proportional to the amount of constraint violations, the suitable value of the penalty factor is chosen after some runs of the algorithm \cite{immanuel}.
According to the equations (\ref{eq pso v} and \ref{eq pso x}) in every iteration each particle of the swarm updates its position coordinates (dimensions) until the termination condition of the algorithm is met.
\section{Study Cases and Simulation Results}\label{simulation}
For analysis and investigation aim, the PSO is applied to find the optimal power flow for the standard IEEE 30-bus system, besides the implementation of the sensitivity analysis to find the most effective control variables for solving the OPF by PSO but in this case within a reduced space of dimensions. The simulation is performed by MATLAB software, the flowcharts of the main parts of simulation codes can be found in \cite{abuella2012study} .
 \subsection{The Data of The System}
The system is shown in Fig. \ref{fig:30bussystem} and the data of the buses, lines, and generators are given in Appendix (\ref{sec:appA}). It consists of six conventional thermal generators at buses 1, 2, 5, 8, 11, and 13, and 41 branches, four of them are transformers with off-nominal tap ratios in branches 6-9, 6-10, 4-12, and 28-27. In addition, the buses 10, 12, 15, 17, 20, 21, 23, 24, and 29 are equipped with shunt VAR compensators.
The limits of control variables are indicated in table (\ref{table:30controllimits}).
 For the other operating (state) variables such as voltages at load buses, the limits are [0.95-1.1]. The limits of reactive power of generators $Q_G$ and the transmission lines ratings are both indicated in generator data and line data tables respectively in Appendix (\ref{sec:appA}).
\begin{figure}[!ht]
\begin{center}
\includegraphics[width=2.7in]{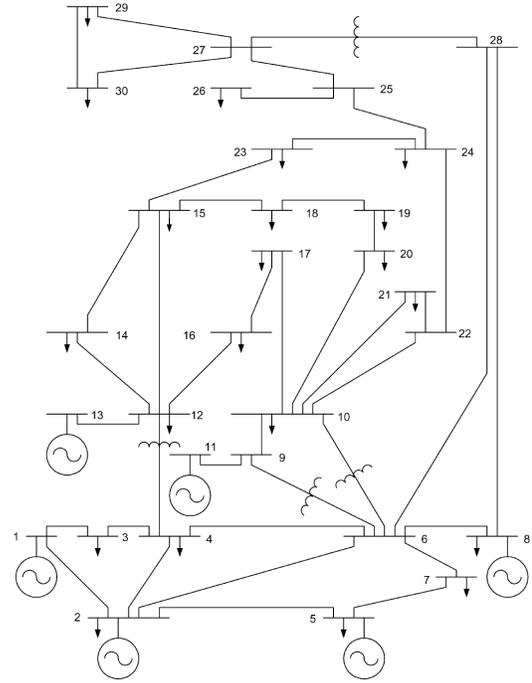}
\caption[Single-line diagram of IEEE 30-bus test system]{Single-line diagram of IEEE 30-bus test system \cite{Abido} \label{fig:30bussystem}}
\end{center}
\end{figure}
\begin{table}[!ht]
\caption{Control variables and their limits\label{table:30controllimits}}
\begin{center}
\includegraphics[width=3.35in]{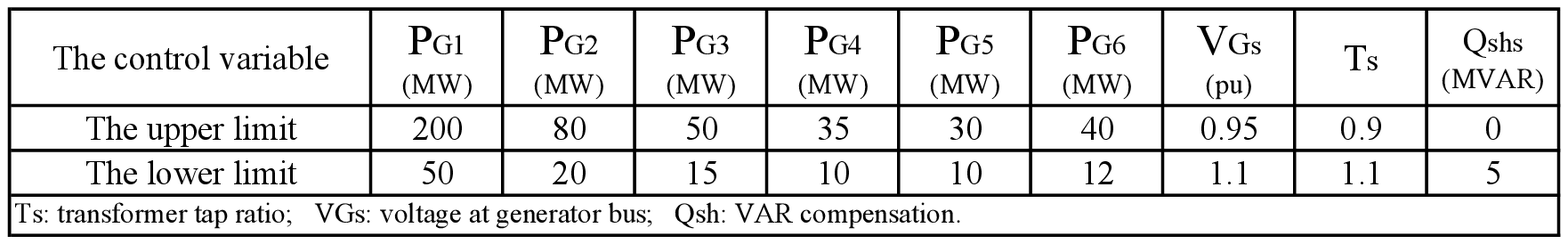}
\end{center}
\end{table}
\subsection{The PSO Algorithm and Its Parameters for Solving OPF}
The skeleton of PSO algorithm is taken form these references \cite{simulator,Brian} after it has been modified for solving OPF. Initially several runs are carried out besides to the helpful information from \cite{Abido,Biskas,Gaing} to select the suitable parameters for PSO algorithm. The inertia weight is decreasing linearly with iterations from its initial value at 0.9 to its ultimate value at 0.4, while the acceleration factors $C_1$ and $C_2$ are equal to 2 and the number of particles is 10. The termination condition occurs when the 5 significant digits after the decimal point of the optimal solution have not changed for last 50 iterations, the algorithm will then consider this as an optimal solution. In addition, the maximum number of iterations after which the algorithm also terminates is 500.

\subsection{The Objective Function}
The objective function is to minimize the operating cost of the system as in equation (\ref{eq psoobjunarg}) which is used in PSO algorithm and its constraints are represented by (\ref{eq mulim}). 

The penalty functions are quadratic penalty functions as those in equation (\ref{eq opfaugj}) and its constraints as (\ref{eq limw}).

\subsection{Study of Base Case}
The running power flow of the initial operating point, which is the base case of loading and it is given in bus data table in Appendix (\ref{sec:appA}), yielding violations in the lower limit of voltage at load buses 18, 19, 20, 21, 22, 23, 24, 25, 26, 27, 29, 30. There is no violation for bus 28 although it is also far from generators, that because this bus is fed by two branches and one of them directly from generator at bus 8. Furthermore, there is a rating violation of the transmission line which connects the buses 1 and 2.

\begin{figure}[!ht]
\begin{center}
\includegraphics[width=2.95in]{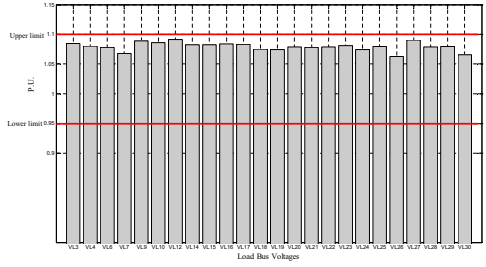}
\caption[Voltage levels at load buses for base case of IEEE 30-bus system]{Voltage levels at load buses for base case of IEEE 30-bus system \label{fig:30basecaseVs}}
\end{center}
\end{figure}

Let's first apply PSO algorithm with all of control variables (i.e. $P_{Gs}, V_{Gs}, T_s$, and $Q_{shs}$) for solving the optimal power flow (OPF) of this base case of loading (283.4$MW$) . For sure, in this case all the violations can be easily alleviated and the voltage at load buses and the transmission lines rating within their limits as they are shown in Fig. \ref{fig:30basecaseVs}. The cost of the real output power of generators is minimized to 798.43 $\$/hr$ as it is illustrated in table (\ref{table:30_283outputs}). \\
By adjusting the voltages at both ends of the transmission line within their limits, the rating constraint of the transmission line 1-2 is alleviated as well for this branch.

\begin{table}[!ht]
\caption{Generators outputs of base case (283.4 $MW$) \label{table:30_283outputs}}
\begin{center}
\includegraphics[width=3.25in]{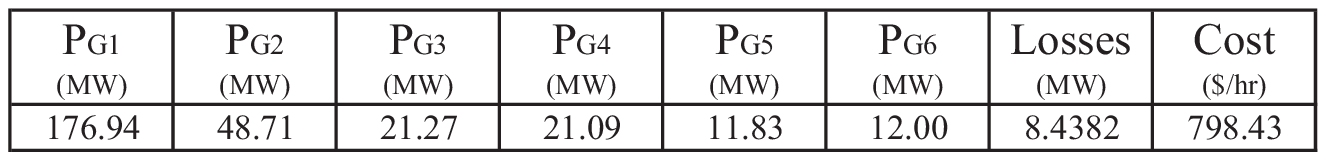}
\end{center}
\end{table}

\subsection{Implementation of Sensitivity Analysis With PSO}
\begin{figure}[!ht]
\centering
\subfloat[Voltage sensitivity matrix $S_u$]{\label{fig:30Su}\includegraphics[width=0.22\textwidth]{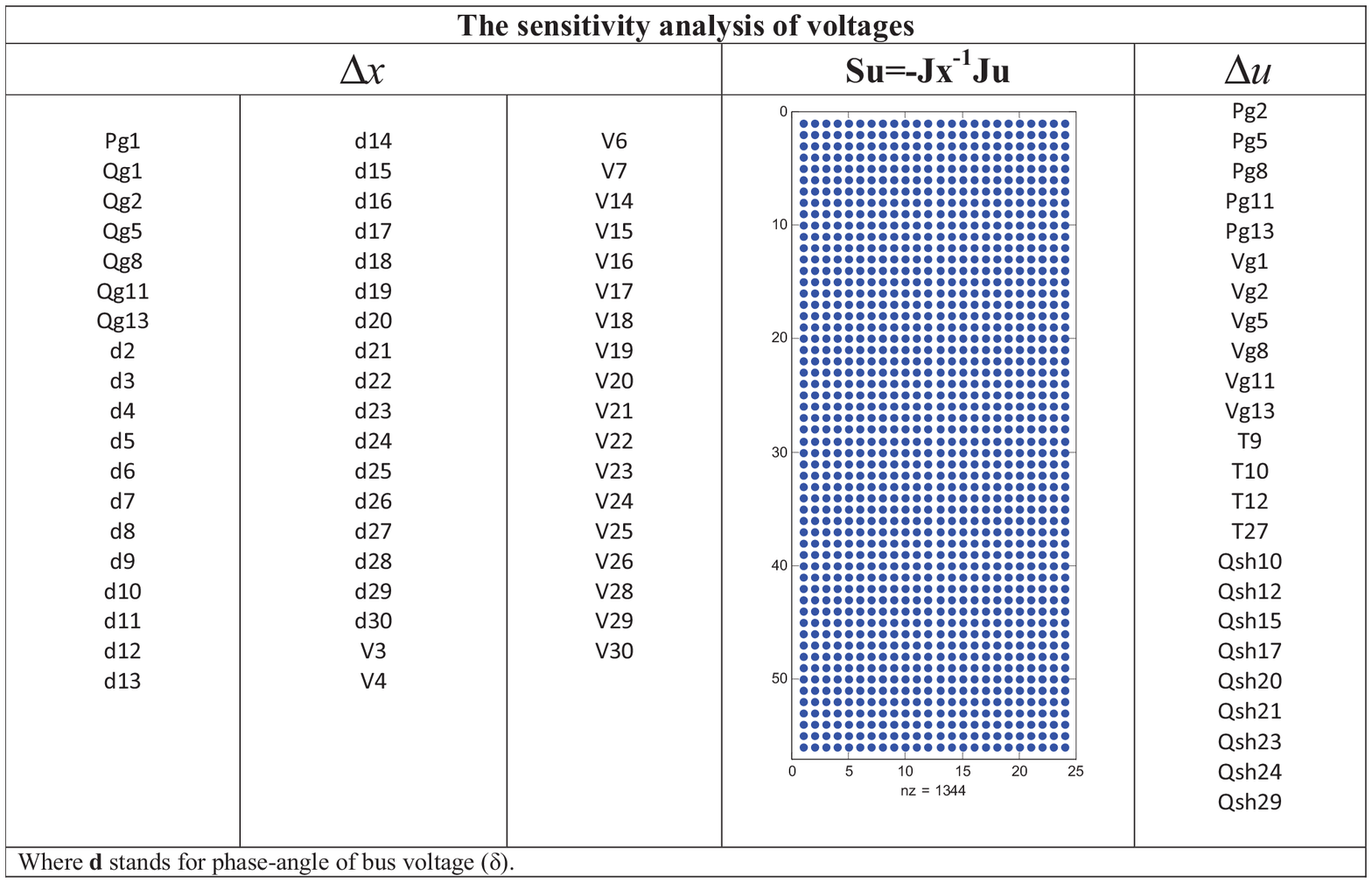}} \ \
\subfloat[Current sensitivity matrix $R$]{\label{fig:30R}\includegraphics[width=0.238\textwidth]{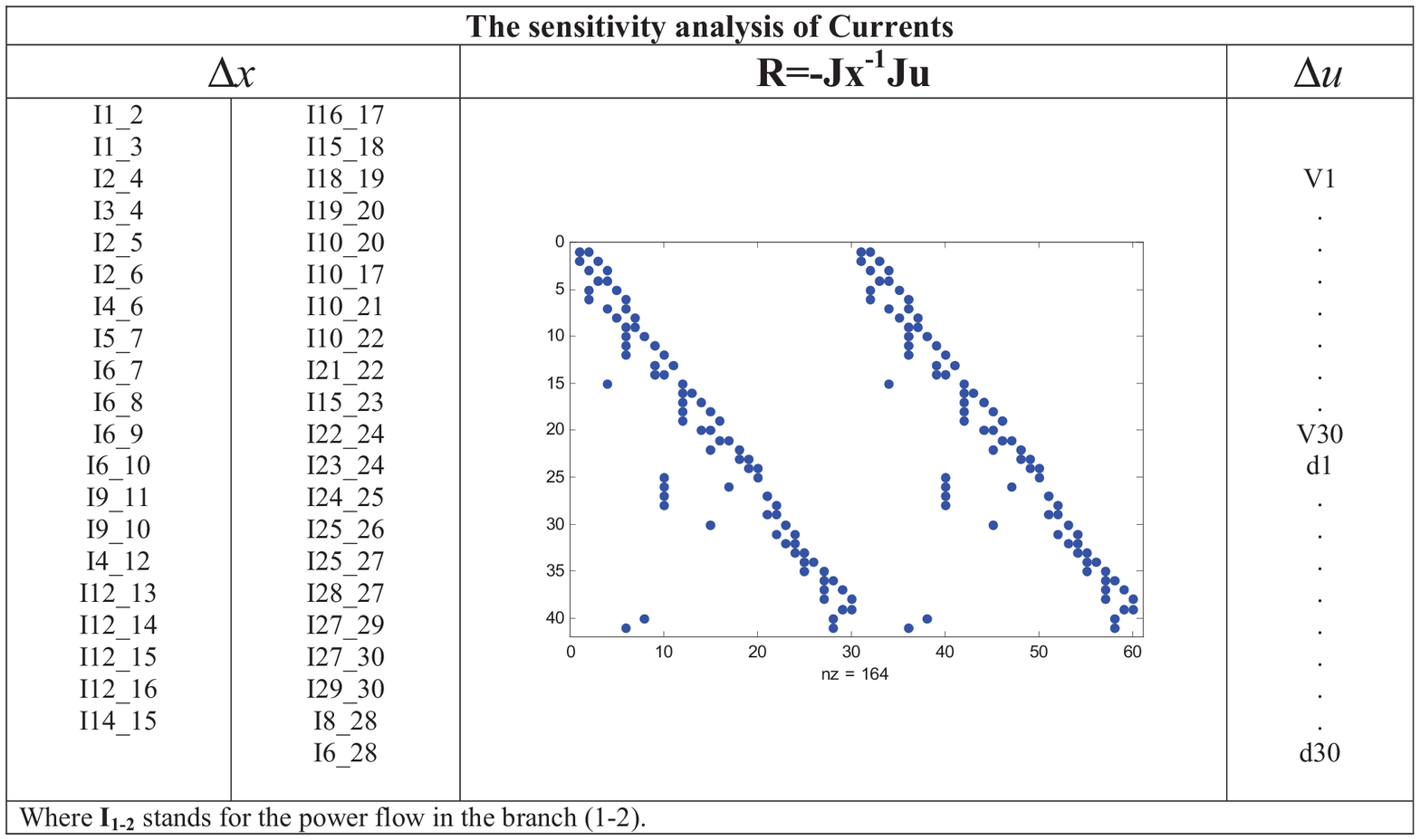}} \\
\hspace{0.3in}\parbox{3.4in}{\caption[Generators' outputs and penalty cost $C_p$ vs. penalty cost,]{Sensitivity matrices and their state $\Delta x$ and control $\Delta u$ variables at base case $283.4 MW$}\label{fig:wckpouts}}\\
\end{figure}

Firstly, the sensitivity analysis of optimal power flow is conducted for the base case to find the sensitivity matrices of the voltage and the current and to select the most effective control variables to adjust the violations in the OPF at the base case within a less-dimensions space of PSO algorithm.
The mathematical approaches of calculating the sensitivity matrices of voltage $Su$ and current $R$ are discussed in (\ref{sec:voltsens} and \ref{sec:cursens}), the resulting rank of state and control variables, in addition to the visualized sparsity pattern of elements in matrices $Su$ and $R$ are shown in Fig. \ref{fig:30Su} and Fig. \ref{fig:30R} respectively.\\

\

Table (\ref{table:voilationsens}) presents the most dominant control variables for every violated bus voltage are ranked from top to down at base case. \\
Notice that the real power of generators $P_{gs}$ are not listed in Table (\ref{table:voilationsens}) because they are also used in the OPF solution search by PSO algorithm to find the minimum cost, although $P_{gs}$ are effectively contributing in adjusting the violations since the constraints are augmented with the main objective function of OPF as in (\ref{eq psoobjunarg}).

This order of control variables are achieved by sensitivity analysis between the voltage sensitivity matrix $S_u$ and its corresponding column vectors of state and control variables. Once the control variables that have most effect on state variables are determined and ranked from the most powerful to the less. Next to that, the selection of these control variable combination is done in PSO for adjusting the violations in operating constraints of the power system by using less number of dimensions.
\begin{table}[!ht]
\caption{Ranked control variables for violated voltages of load buses at Base Case \label{table:voilationsens}}
\begin{center}
\includegraphics[width=3.1in]{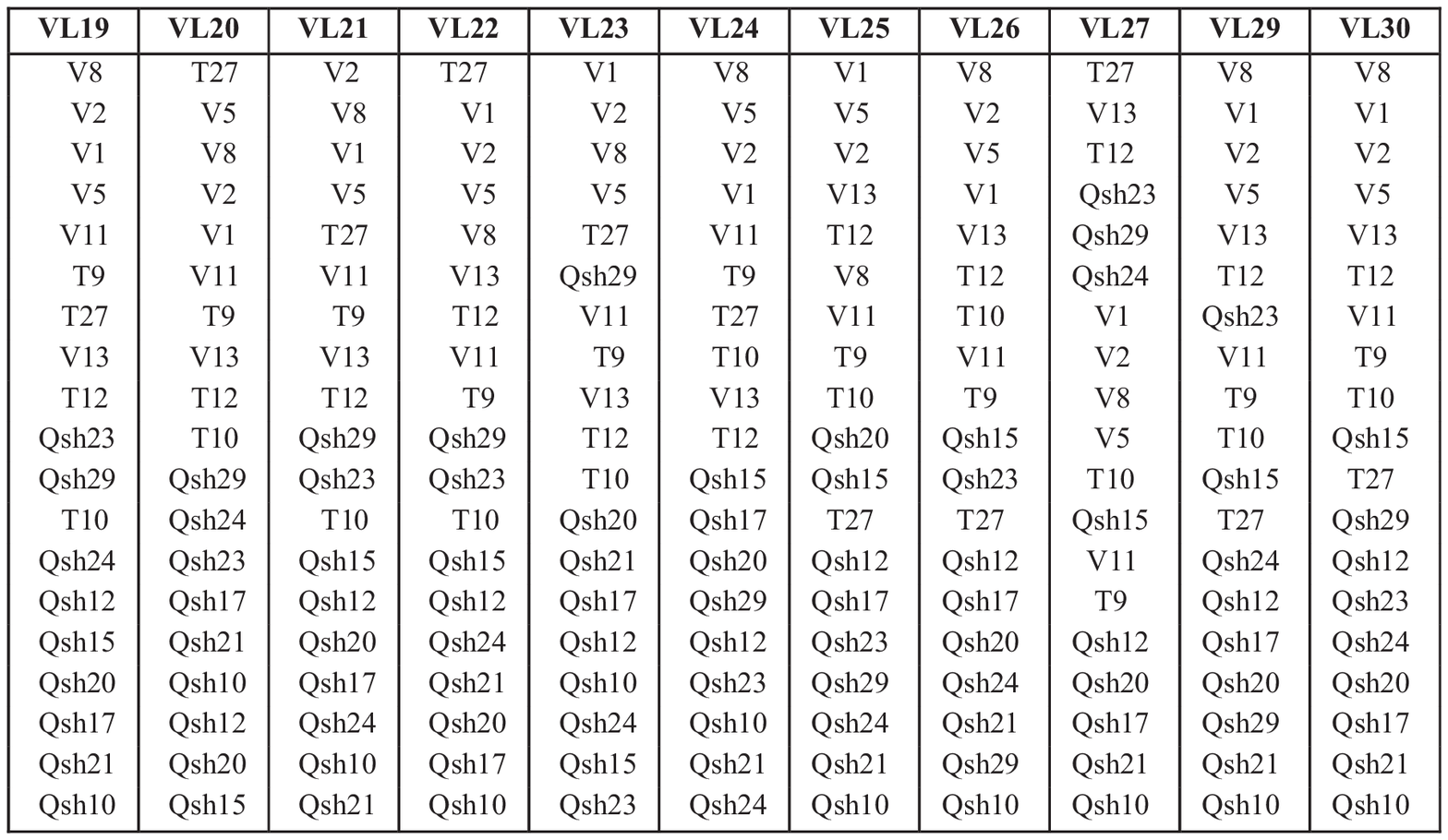}
\end{center}
\end{table}
When control variables increase the dimensions of the particles in PSO algorithm also increase and that can cause a complexity for manipulating the dimensions of the particles \cite{Biskas}. Thereby, finding and using the most effective control variables to adjust and correct the violations can decrease the dimensions in PSO and enhance its performance.
\subsection{PSO Solution for Base Case Using Most Effective Control Variables }
Now let's using only the most effective control variables to adjust the violations in initial operating point of IEEE 30-bus system.
Several combinations of dominant control variables (as they are ranked in table (\ref{table:voilationsens})) can be chosen, some of them are sufficient combination to bring back the violations in the voltage at load buses within their limits for base case of 283.4 $MW$ loading level. The PSO's results with the combinations of most effective control variables are listed in table (\ref{table:2834controlcases}).
\begin{table}[!ht]
\caption{PSO result of combinations of most effective control variables at base case \label{table:2834controlcases}}
\begin{center}
\includegraphics[width=3.4in]{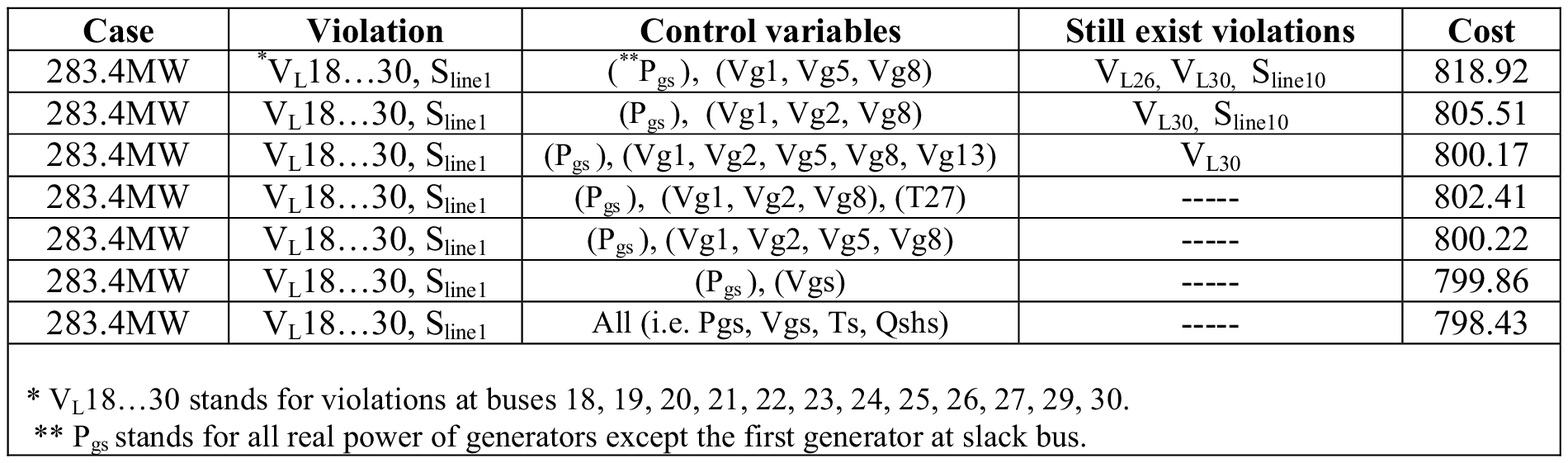}
\end{center}
\end{table}
As it is shown in table (\ref{table:2834controlcases}), except the case of using all control variables, it's obvious that the control variables combination of $ P_{gs}$ and $V_{gs}$ is the most effective with minimized cost 799.86 $\$/hr$. While for other less number of control variables, the combination of most effective control variables of $P_{gs}, V_{g1}, V_{g2}, V_{g5}$ and $V_{g8}$ and the other of $P_{gs}, V_{g1}, V_{g2}, V_{g8}$ and $T_{27}$ succeed to adjust all violations, but the former combination produces lower cost 800.22 $\$/hr$.

\subsection{PSO Solution for Different Loading Levels}
For IEEE 30-bus test system with other cases of loading, higher and lower than the base case, PSO is used with only the most effective control variables to adjust the violations, if they exist. The results are shown in table (\ref{table:Flscases}).
\begin{table}[!ht]
\caption{PSO result for several loading cases of IEEE 30-bus test system \label{table:Flscases}}
\begin{center}
\includegraphics[width=3.25in]{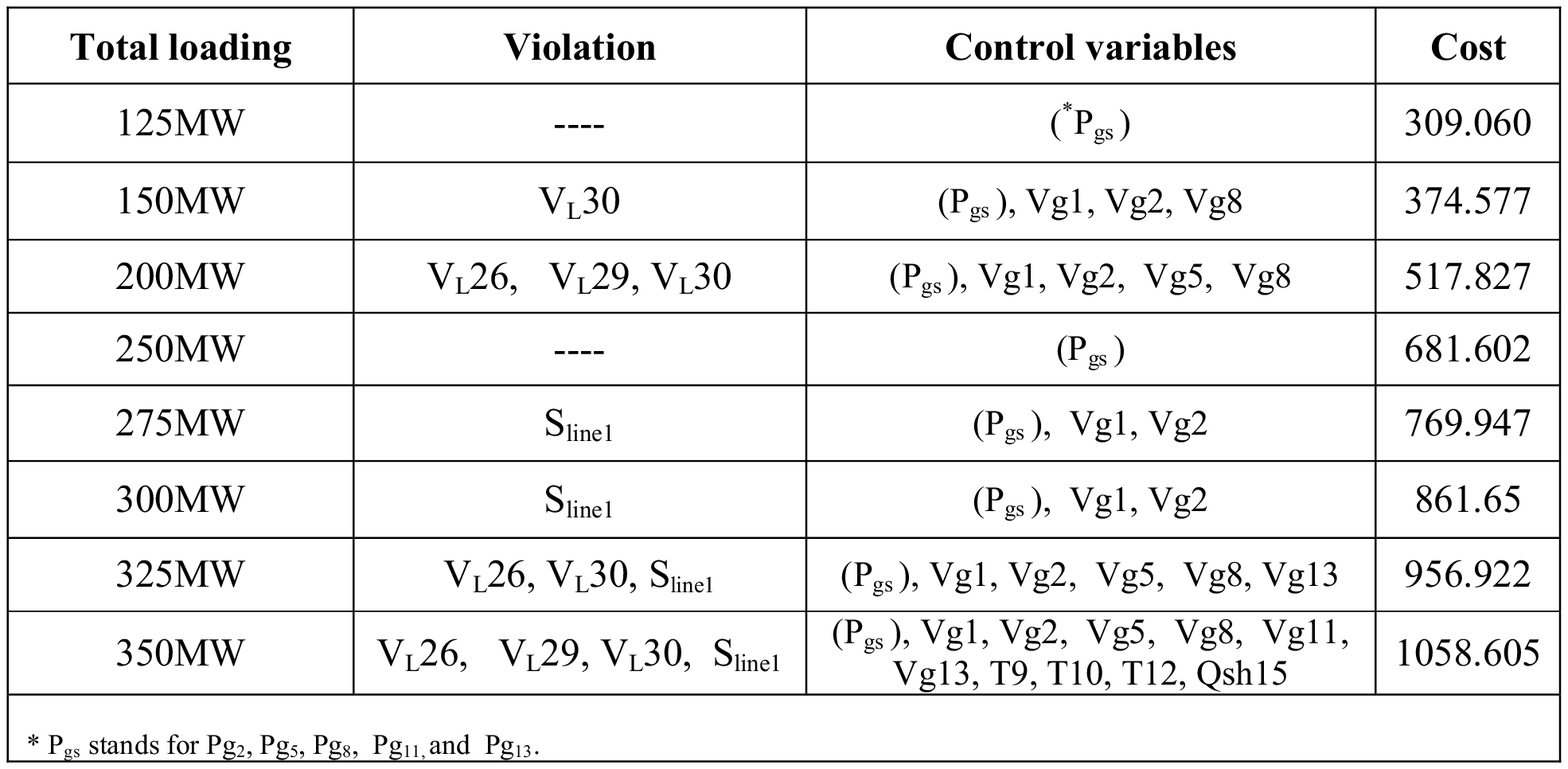}
\end{center}
\end{table}

Note: Every OPF solution of the previous loading case is considered as an initial operating point for the following loading case. for example, if the OPF of the loading case 125 $MW$ has been solved then the OPF of the successive loading case 150 $MW$ considers the solution of the previous loading case 125 $MW$ as its initial point and so on.\\
$V_{L26}, V_{L29}$, and $V_{L30}$ are the weakest buses in the system that are susceptible to violations more than other buses. While the transmission line that connect bus 1 and bus 2 is the weakest transmission line and it suffers from violation of its rating for several loading cases. In the last case 350 $MW$ a variety of control variables are needed to keep the system in secure operation, but they are still less than using all control variables.
\begin{figure}[!ht]
\centering
\subfloat[150 $MW$]{\label{fig:150MW}\includegraphics[width=0.25\textwidth]{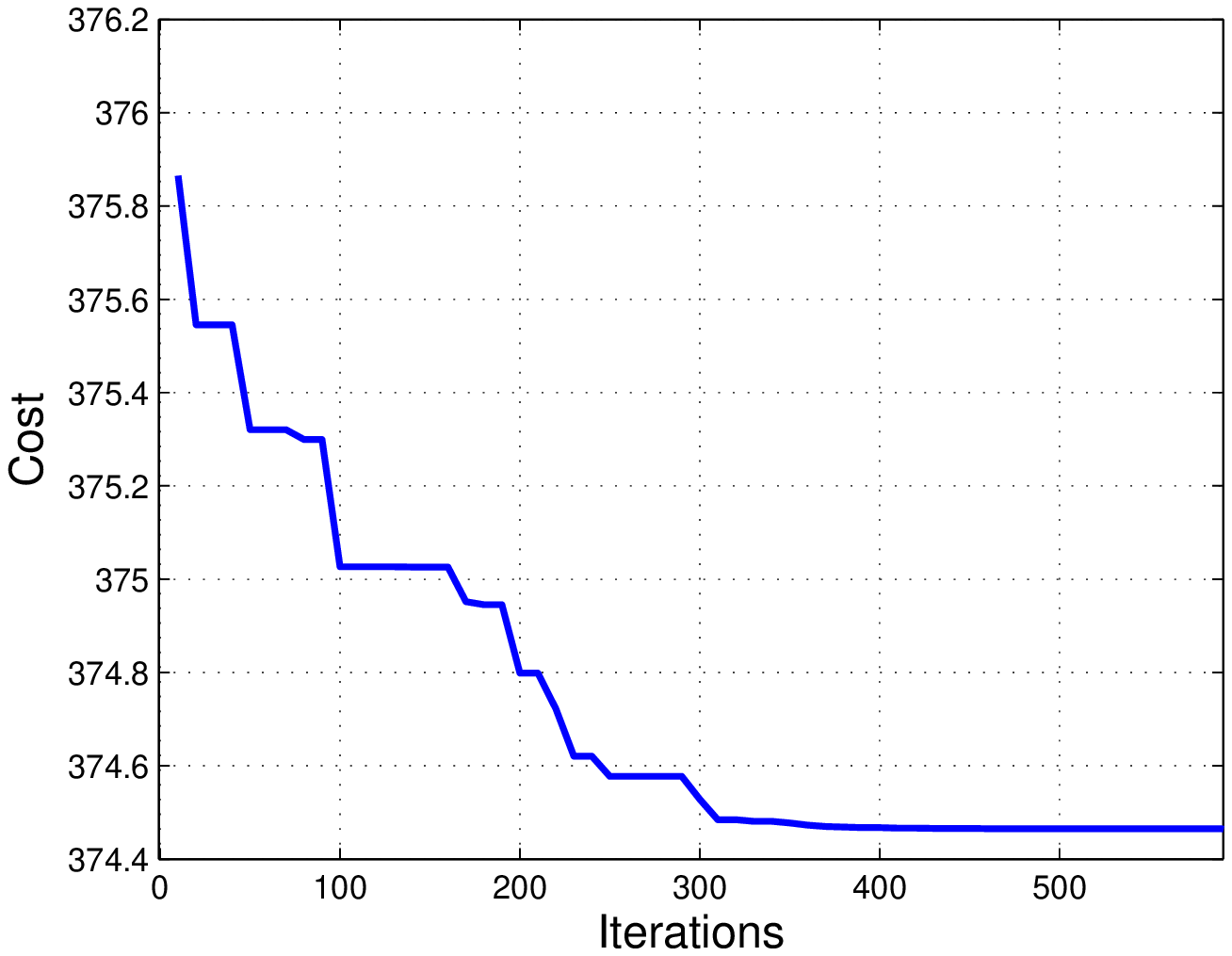}}
\subfloat[325 $MW$]{\label{fig:325MW}\includegraphics[width=0.25\textwidth]{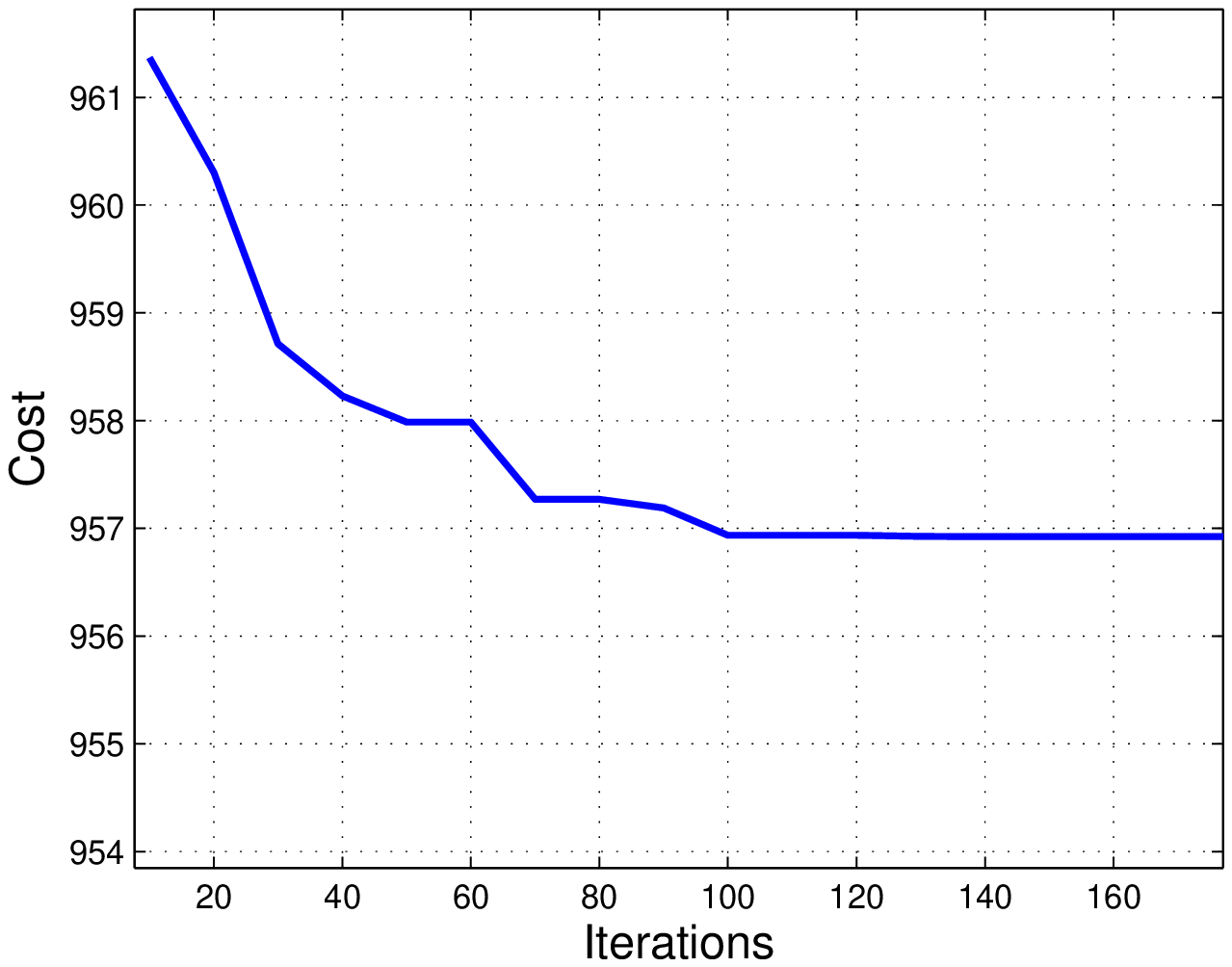}}
\caption{Cost vs. iterations of two loading cases}
\label{fig:FLcases}
\end{figure}\
\section{Conclusion}  \label{conclusion}\
The implementation of PSO algorithm to solve the OPF problem is useful and worth of investigation. Moreover, PSO algorithm is easy to apply and simple since it has fewer number of parameters to deal with, comparing to other modern optimization algorithms.
In addition, PSO algorithm is appropriate for solving the optimal power flow for systems that include variable generation resources. 
Using most effective control variables by applying sensitivity analysis reducing the space dimensions of PSO and hence improving the computing effort is needed for PSO algorithm and enhancing its performance, especially for large systems including many stochastic generation resources.

 The following could be included for further work: PSO algorithm needs some work on selecting proper parameters and it also needs more accurate mathematical description for its convergence. PSO can be applied in wind power bid marketing between electric power operators. In addition to operating cost, the environment effects and security or risk of wind power penetration can be included by using multi-objective models. Using singular value decomposition and pseudo-inverse techniques could be considered for further study to find the effective control variables. \\

\appendix 


\subsection{The Data for IEEE 30-Bus Test System}\label{sec:appA}

The data for IEEE 30-bus test system as following \cite{stott}:\
\begin{table}[!ht]
\caption{Bus data of IEEE 30-bus test system\label{table:30busdata}}
\begin{center}
\includegraphics[width=3.in]{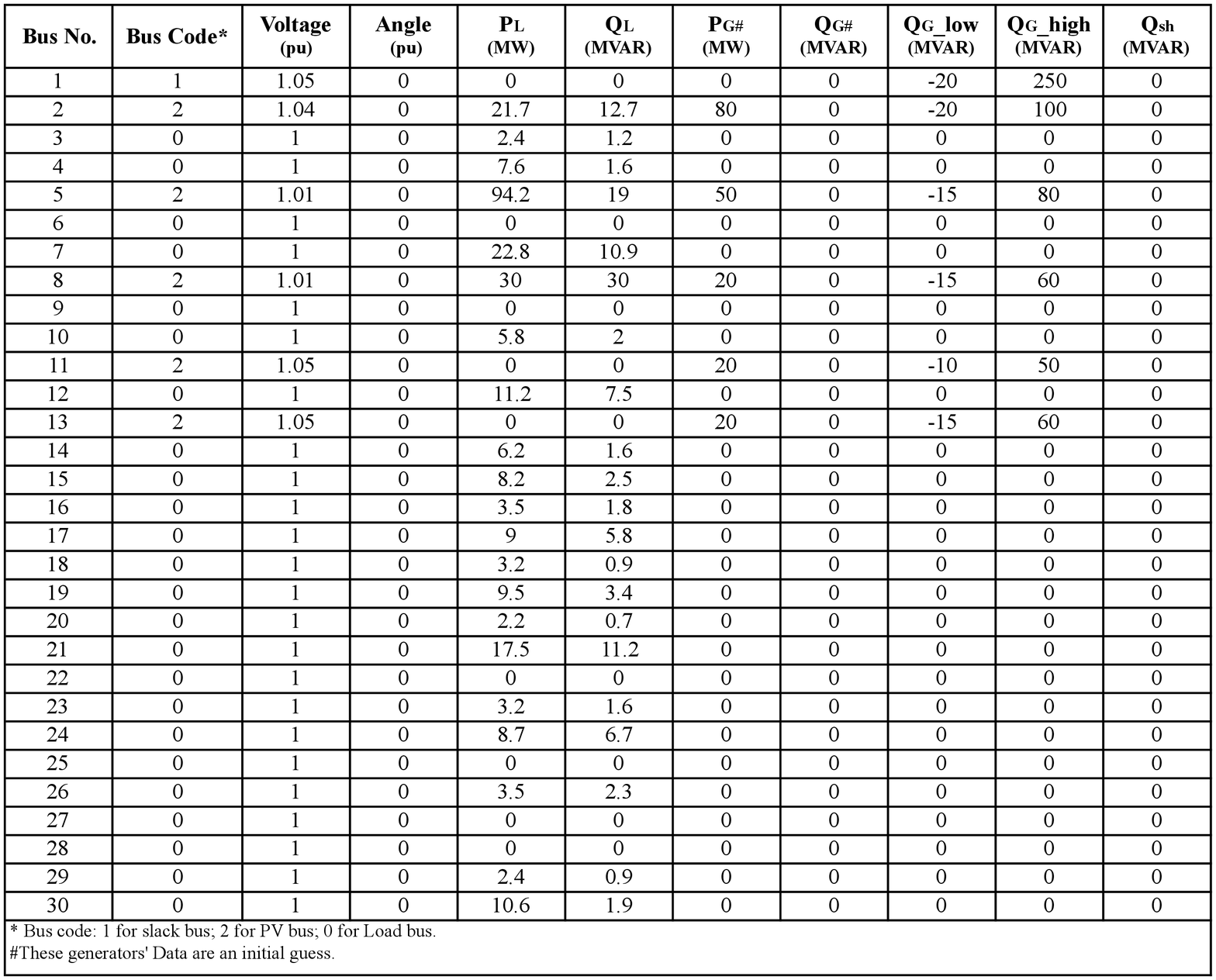}
\end{center}
\end{table}\\\
\begin{table}[!ht]
\caption{Generators data of IEEE 30-bus test system\label{table:30gendata}}
\begin{center}
\includegraphics[width=3in]{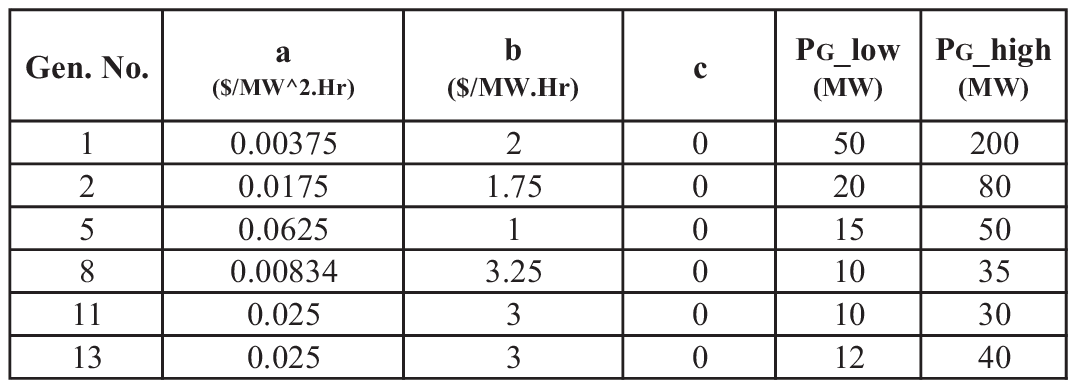}
\end{center}
\end{table}\
\begin{table}[!ht]
\caption{Line data of IEEE 30-bus test system\label{table:30linedata}}
\begin{center}
\includegraphics[width=3in]{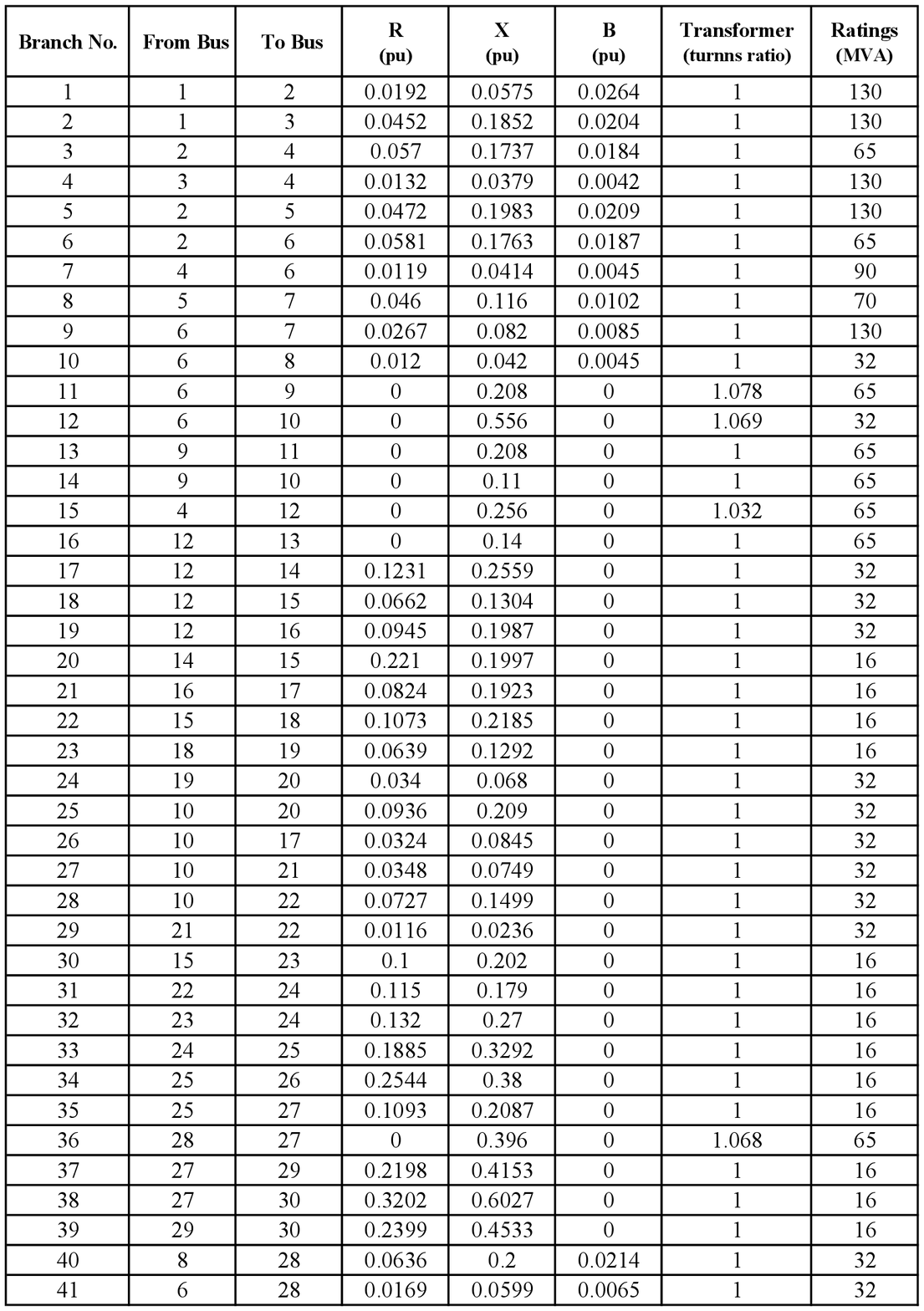}
\end{center}
\end{table}\\


\section{References}
\bibliographystyle{ieeetr}

\bibliography{Refs1}

\section*{biographies}

\footnotesize{\textbf{Mohamed A. Abuella} (IEEE student member). He is a PhD student at University of North Carolina at Charlotte. He received his Bachelor's degree from College of industrial Technology, Misurata, Libya in 2008, and MSc degree from Southern Illinois University Carbondale in 2012. His research interests are in planning and operations of power systems.}\\

\footnotesize{\textbf{C.J. Hatziadoniu} (IEEE M'87, SM'06). He is a professor of electrical and computer engineering at Southern Illinois University Carbondale. His research interests include power systems control and protection and application of power electronics to power systems.}

\end{document}